\documentclass[prl,superscriptaddress]{revtex4}
\usepackage{graphicx}

\begin{document}

\title{Giant paramagnetic Meissner effect in multiband superconductors}

\author{R. M. da Silva}
\affiliation{Programa de P\' os-Gradua\c c\~ao em Ci\^encia dos Materiais, Universidade Federal de Pernambuco, Av. Jorn. An\'{i}bal Fernandes,
s/n, 50670-901 Recife-PE, Brazil}
\author{M. V. Milo\v{s}evi\'c}
\affiliation{Departement Fysica, Universiteit Antwerpen,
Groenenborgerlaan 171, B-2020 Antwerpen, Belgium}
\author{A. A. Shanenko}
\affiliation{Departamento de F\'isica, Universidade Federal de Pernambuco, Av. Jorn. An\'{i}bal Fernandes, s/n,  50670-901 Recife-PE, Brazil}
\author{F. M. Peeters}
\affiliation{Departement Fysica, Universiteit Antwerpen, Groenenborgerlaan 171, B-2020 Antwerpen, Belgium}
\author{J. Albino Aguiar}
\affiliation{Departamento de F\'isica, Universidade Federal de Pernambuco, Av. Jorn. An\'{i}bal Fernandes, s/n,  50670-901 Recife-PE, Brazil}
\affiliation{Programa de P\' os-Gradua\c c\~ao em Ci\^encia dos Materiais, Universidade Federal de Pernambuco, Av. Jorn. An\'{i}bal Fernandes,
s/n, 50670-901 Recife-PE, Brazil}
\date{\today}
\begin{abstract}
Superconductors, ideally diamagnetic when in the Meissner state, can also exhibit paramagnetic behavior due to trapped magnetic flux. In the absence of pinning such paramagnetic response is weak, and ceases with increasing sample thickness. Here we show that in multiband superconductors paramagnetic response can be observed even in slab geometries, and can be far larger than any previous estimate - even multiply larger than the diamagnetic Meissner response for the same applied magnetic field. We link the appearance of this {\it giant paramagnetic response} to the broad crossover between conventional Type-I and Type-II superconductors, where Abrikosov vortices interact non-monotonically and multibody effects become important, causing unique flux configurations and their locking in the presence of surfaces.
\end{abstract}
\maketitle

\section*{Introduction}
The diamagnetic Meissner effect is one of the hallmarks of superconductivity, where applied magnetic field is ideally screened out of the superconductor when cooled below the critical temperature $T_c$. However, many field-cooled experiments on various materials over the past two decades detected a paramagnetic response, i.e. {\it enhanced} magnetic field inside the sample, usually referred to as paramagnetic Meissner effect (PME) or Wohlleben effect. The materials in question range from elementary ones such as Nb \cite{NbPMEobs1,NbPMEobs2}, to much more complex high-$T_c$ cuprates \cite{HTSPMEobs1,HTSPMEobs2,HTSPMEobs3,HTSPMEobs4,HTSPMEobs4-2,HTSPMEobs5,HTSPMEobs6}. One proposed explanation for the enigmatic origin of PME in cuprates is based on the $d$-wave symmetry of the order parameter and the idea that $\pi$ junctions formed due to Josephson coupling between grain boundaries can result in spontaneous current loops with significant magnetic moments \cite{omm1,omm1-2,omm2,omm3,omm4,omm5}. A much simpler and more general explanation is the compression and trapping of magnetic flux on cooling. Using this picture, Koshelev and Larkin \cite{kosh} calculated the magnitude of PME in thin stripes of conventional superconductors, and concluded that its theoretical maximum is $\sim27\%$ of the full Meissner response for the given magnetic field. Kosti\'{c} {\it et al.} \cite{kost} performed a supporting experiment on bulk Nb, and further concluded that polishing the sample surfaces strongly alters the PME, i.e. that surface barriers for flux entry and exit play an important role.

The appearance of PME due to flux compression is easiest to understand in the case of mesoscopic samples, where the influence of confining boundaries is crucial. There, in analogy to surface superconductivity, during field-cooling the superconducting order parameter nucleates at the sample surface, and traps a multiquanta (giant) vortex inside the sample. Such large and compressed flux may lead to paramagnetic response, as first predicted by Moshchalkov {\it et al.} using self-consistent Ginzburg-Landau simulations \cite{mosh}, and subsequently verified experimentally by Geim {\it et al.} \cite{geim}. A simple consideration shows that the paramagnetic moment strongly depends on the sample thickness, so that in very thick samples it appears only at very large fields and scales with penetration depth $\lambda$ over lateral size of the sample, while in thin plates it can be significant and scales with $\lambda$ over thickness \cite{dSSilva2001,Barba2008}. Therefore, the enigmatic PME in high-temperature superconductors becomes intrinsic for thin mesoscopic conventional superconductors.

Recent years have seen the rise of interest in {\it multiband superconductivity}, particularly since its discovery in MgB$_2$ and in iron-based materials \cite{mult1,mult2,mult3}. The former has the highest $T_c$ of intermetallics; the latter are layered and high-temperature superconductors. To date, there have been no investigations of the paramagnetic response in these materials. Instead, a lot of attention has been paid to their rich vortex matter \cite{baba}, and their possible classification outside the Type-I/Type-II dichotomy \cite{moshch,rev_brandt} due to observed non-monotonic vortex interaction \cite{Chaves2011}. Early works on single-band superconductors already discussed the broad crossover between conventional types of superconductivity  \cite{t21_1,t21_2,t21_3,t21_4,t21_5,t21_5-2,t21_6}, where Abrikosov vortices exhibit long range attraction and penetration of vortices manifests as a large magnetization jump from the Meissner state to the mixed state (see e.g. Ref. \cite{t21_4}). The lower bound of the crossover is given by the $H_c(T)=H_{c2}(T)$ line in the parametric space ($H_c$ being the thermodynamic critical field and $H_{c2}$ the upper critical field) \cite{t21_4,t21_5,t21_5-2,t21_6,t21_7,t21_8}, below which textbook Type-I behavior takes place (for $H_c>H_{c2}$ only Meissner state is thermodynamically stable, unless mesoscopic effects are strong, see Ref. \cite{LukVin}). The disappearance of the long-range vortex attraction marks the end of the crossover domain, and conventional Type-II behavior is recovered. This picture was recently detailed and extended to the multiband case in Ref. \cite{vagov}. It is clear that non-monotonic vortex interaction and other interplay effects between condensates in multiband superconductors are bound to also affect the interactions of trapped magnetic flux with the sample boundaries, and can lead to novel manifestations of the paramagnetic Meissner response. To reveal, quantify and explain the latter is the core objective of the present report.

\section*{Results}
\begin{figure}[t]
\includegraphics[width=0.4\linewidth]{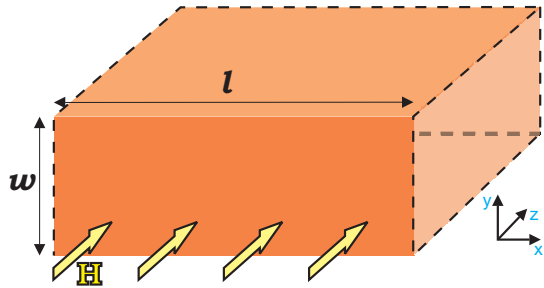}
\caption{ Oblique view of the sample, the superconducting slab of width $w$, very long in other dimensions (indicated by dashed lines), in parallel magnetic field $H$.} \label{scheme}
\end{figure}

We consider a larger than mesoscopic two-band superconducting slab of width $w$ ($w/\lambda$ ranges from $15$ to $50$ for the considered parameters) in a parallel magnetic field (see Fig. \ref{scheme}), and report particular behavior of the sample magnetization as a function of the applied field [M(H) loops]. We primarily focus on two-band materials, but our findings can be qualitatively extrapolated to systems with more than two bands. The calculations are performed within the two-component Ginzburg-Landau (TCGL) theory (see Methods), where we have cautiously set a sufficiently high temperature $T$ to ensure the qualitative and quantitative validity of
our predictions in the context of recent debates \cite{Geilik1967,Kogan2011,comBab2012,Shanenko2011,Vagov2012,Silaev2012}, and we used full microscopic expressions of all coefficients in the theory \cite{Zhitomirsky2004,Vagov2012,Silaev2012}. TCGL theory then comprises eight independent parameters, namely, the Fermi velocities of the bands $v_1$ and $v_2$, the elements of the coupling matrix $\lambda_{11}$, $\lambda_{22}$ and $\lambda_{12}=\lambda_{21}$, the
total density of states $N(0)$ as well as the partial density of states of the first band $n_1$ (note $n_1+n_2=1$), and finally $T_{c}$, which sets the energy scale $W^2=8\pi^2T_c^2/7\zeta(3)$. By fixing the unit of length $\zeta_1$ and normalizing the order parameters by $W$, the parameters $v_1$ and $T_c$ are fixed, and we are left with six parameters in the model: $\lambda_{11}$, $\lambda_{22}$, $\lambda_{12}$, $v_1/v_2$, $n_1$ and $N(0)$. Instead of choosing $N(0)$, we opt to show the GL parameter of the first (stronger) band-condensate  $\kappa_1=\frac{3cW}{hev_1^2}\sqrt{\frac{\pi}{2n_1N(0)}}$, which is an indicator for the expected magnetic behavior of the sample.


In what follows, we consider an infinitely thick slab of width $w=80\zeta_{1}$, and use periodic boundary conditions in the longitudinal direction (with size of the unit cell $l=120\zeta_1$, see Fig. \ref{scheme}).
Without loss of generality, we take for the remaining microscopic parameters of the sample: $\kappa_1=1.5$, $\lambda_{11}=1.55$, $\lambda_{22}=1.3$, $\lambda_{12}=0.09$ and $n_1=0.48$. Note that such choice of parameters does not correspond to any particular material, and is actually by no means unique - since our main study will concern the dependence of the magnetic properties on the ratio of the Fermi velocities $v_1/v_2$ and temperature. We demonstrate these properties via calculated magnetization [$M(H)$] loops while adiabatically sweeping the magnetic field up and down.
 
\begin{figure}[t]
\includegraphics[width=\linewidth]{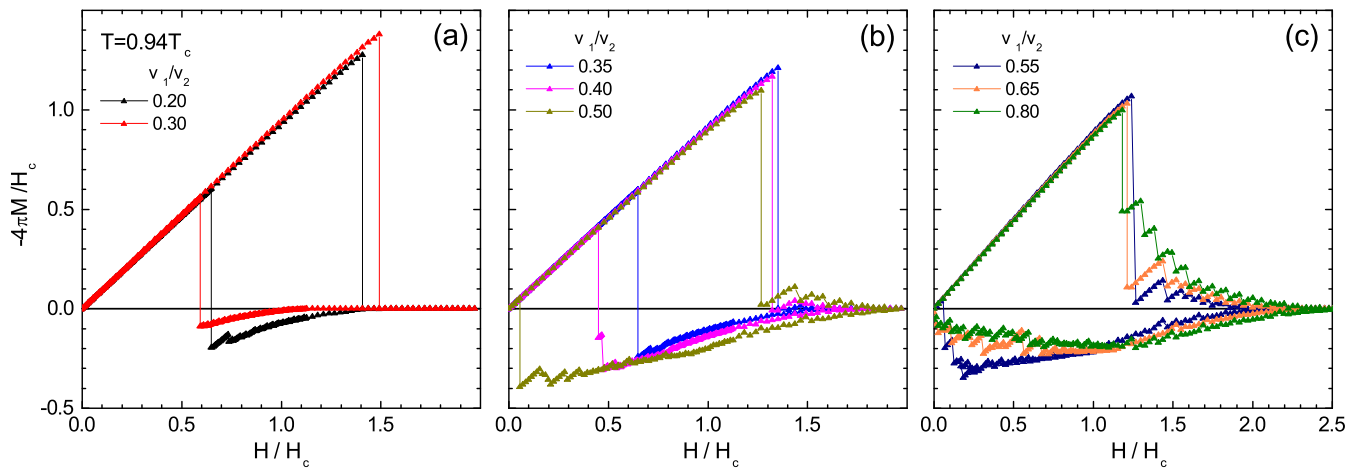}
\caption{ Magnetization $M(H)$ loops at
$T=0.94T_c$, for sequentially increased ratio of the Fermi velocities $v_1/v_2$ (and other parameters $\lambda_{11}=1.55$, $\lambda_{22}=1.3$, $\lambda_{12}=0.09$, $n_1=0.48$, and $\kappa_1=1.5$), obtained by sweeping up and down the external magnetic field $H$ (given in units of the thermodynamic critical field $H_c$).} \label{mag-loop}
\end{figure}

In Fig. \ref{mag-loop} we show the $M(H)$ loops at $T=0.94T_c$, for different values of $v_1/v_2$. By increasing the latter parameter, we are actually decreasing the characteristic length scale of the second condensate $\zeta_{2}=\hbar v_2/\sqrt{6}W$ (since $\zeta_1$ is fixed as the unit of distance) and we are thereby driving the system into the Type-II magnetic behavior (since $\kappa_2=\kappa_1\frac{v_1^2}{v_2^2}\sqrt{\frac{n_1}{n_2}}$ and the GL parameter of the coupled system $\kappa$ \cite{Kogan2011} are increasing; for calculation of the penetration depth $\lambda$, please see Ref. \onlinecite{Chaves2011}). This directly manifests in magnetization curves: for low $v_1/v_2(\lesssim0.3)$ one easily recognizes typical response of a Type-I slab (see Fig. \ref{mag-loop}(a)), with superheated Meissner state in increasing field (with subsequent collapse to normal state), and supercooling in decreasing field \cite{Andre2012}, with some flux trapping present; for high $v_1/v_2(\gtrsim 0.65)$, the expected response of a Type-II slab is recovered \cite{Clem1974}, still with some paramagnetic flux trapping (see Fig. \ref{mag-loop}(c)). However, at intermediate values of $v_1/v_2$ {\it a uniquely different shape of the magnetization loop is found}, with a pronounced jump from the Meissner state to the mixed state with increasing field, and a very pronounced paramagnetic response in decreasing field (see Fig. \ref{mag-loop}(b)).

\begin{figure}[t]
\includegraphics[width=0.6\columnwidth]{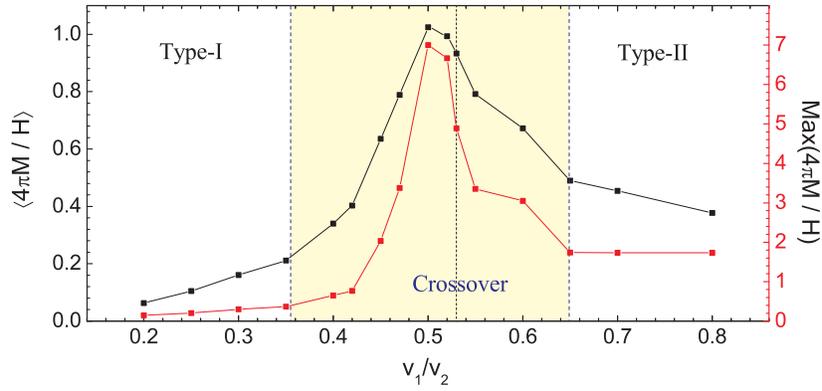}
\caption{Maximal paramagnetic response in decreasing field at $T=0.94T_c$ (red) and its total cumulative value over the field span (black), as a function of $v_1/v_2$. Vertical lines indicate where $H_c=H_{c2}$, where the S-N surface energy changes sign (i.e. $\sigma_{SN}=0$), and where long-range interaction of vortices changes sign (left to right, respectively), delimiting the crossover range between standard types of superconductivity.} \label{ParResp}
\end{figure}

In increasing field, all calculated magnetization loops exhibit a superheated Meissner state above the thermodynamic critical field $H_c$, where the superheating field $H_{sh}$ agrees very well with the seminal calculations of Matricon and Saint James for $H_{sh}(\kappa)$ of single-band materials \cite{Matricon1967}. At $H=H_{sh}$, superconductivity is either destroyed (for $v_1/v_2<0.34$) or a jump to the mixed state occurs (for $v_1/v_2>0.34$). The delimiting value of $v_1/v_2=0.34$ exactly satisfies the condition $H_c=H_{c2}$. In decreasing magnetic field, the superconductivity nucleates at the surface superconductivity field $H_{c3}$ \cite{Zhitomirsky2004}. Indeed, the nucleated states were only superconducting at the surfaces of the slab, with a large normal domain in the interior of the slab. For further lowered field and $v_1/v_2<0.34$ the normal domain remains trapped until abruptly expelled from the sample at the expulsion field $H_e$. This analysis confirms that magnetic response of the system for $v_1/v_2<0.34$ is the one of Type-I superconductors, since typical superheating-supercooling picture holds there, $H_{c2}$ is smaller than $H_c$, and no vortices are found in the paramagnetic branch where flux was trapped upon nucleation of surface superconductivity. However, while decreasing field for $v_1/v_2>0.34$, where consequently $H_{c2}>H_c$, the normal domain becomes unstable at field $H_d$ but is not expelled; instead, it spreads into a vortex configuration, stable down to persistently lower expulsion field $H_e$ as $v_1/v_2$ is increased. Simultaneously, flux trapping becomes notably more efficient, so that the vortex exit is hampered in decreasing field and paramagnetic response increases to its maximum at $H_e$. This tendency continues up to $v_1/v_2\approx 0.53$, for which paramagnetic response is almost an order of magnitude larger than the Meissner response at $H=H_e$, and approximately 30 times larger than the largest theoretical estimate of paramagnetic response to date (scaled to the diamagnetic response at a given field, see Ref. \cite{kosh}). We therefore refer to this property as {\it giant paramagnetic response} (GPR). For $v_1/v_2>0.53$, the cumulative paramagnetic response is still very large but gradually decreases, and magnetization curves in decreasing field connect to zero without any abrupt flux expulsion. In other words, we approach the Type-II limit, in which magnetization is expected to hover around zero for descending field in the presence of surface barriers \cite{Clem1974}. In Fig. \ref{ParResp}, we summarize the observed maximal amplitude, $\textrm{Max}(4\pi M / H)$ in the entire field range, and the total cumulative paramagnetic response, $\langle 4\pi M/H\rangle = \frac{4 \pi}{H_c} \int_{H_c}^{0}(M/H)dH$, as a function of $v_1/v_2$, extracted from Fig. \ref{mag-loop}.

\begin{figure}[t]
\includegraphics[width=0.6\columnwidth]{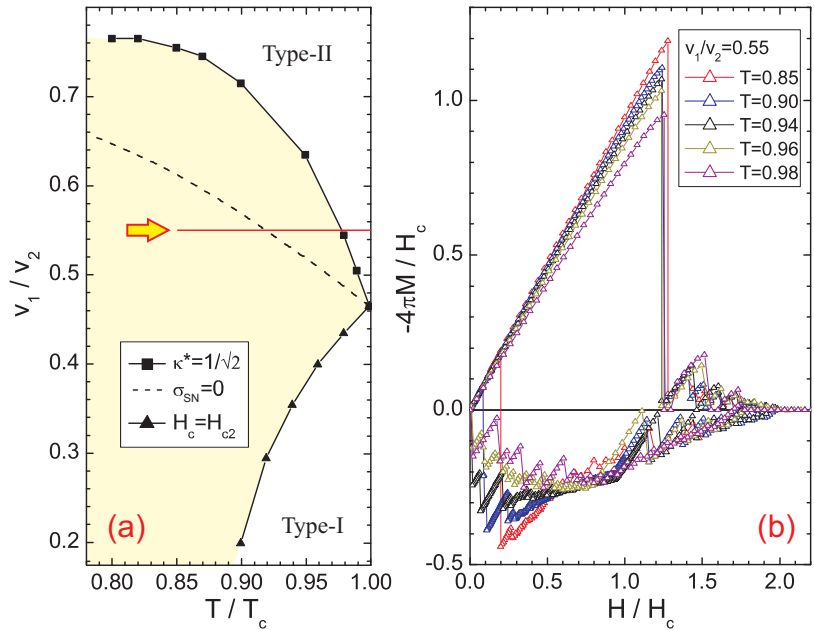}
\caption{(a) The boundaries between different types of superconductivity in the $(v_1/v_2,T)$ plane, for other parameters as in Fig. \ref{mag-loop}. $\kappa^*=1/\sqrt{2}$ line marks the onset of long-range attraction between vortices. At $H_c(T)=H_{c2}(T)$ line, the mixed state vanishes in the bulk material. Dashed line shows where the energy of the superconductor-normal metal interface ($\sigma_{SN}$) changes sign. Arrow shows the path to obtain the sequence of magnetization curves shown in panel (b), for $v_1/v_2=0.55$ and varied temperature. Distinct changes in $M(H)$ loops are found when either curve in panel (a) is crossed.} \label{critdom}
\end{figure}

Based on Fig. \ref{ParResp}, we argue that the giant paramagnetic response is characteristic for superconductors between conventional Type-I and Type-II \cite{vagov}. Namely, this pronounced paramagnetic response is exactly found for sample parameters between the line $H_c(T)=H_{c2}(T)$ and the line where long-range vortex interaction changes sign (determined by effective GL parameter $\kappa^*$ calculated after Ref. \cite{Chaves2011}), with a maximum found close to the parametric line where surface energy ($\sigma_{SN}$) of the superconductor-normal metal (S-N) interface changes sign (determining the change in the polarity of the short-range vortex interaction \cite{Chaves2011}). For the microscopic parameters considered here, we show this domain in Fig. \ref{critdom}(a), as a function of $v_1/v_2$ and temperature. 
To test our hypothesis further, we calculated an additional set of $M(H)$ loops, shown in Fig. \ref{critdom}(b), for fixed $v_1/v_2 = 0.55$ and varied temperature indicated by yellow arrow in Fig. \ref{critdom}(a). From Fig. \ref{critdom}(b),  we confirmed exactly the same behavior of the loops and relationship of the giant paramagnetic response (GPR) with the delimiting lines of the critical domain: for $T \geq 0.98 T_c$ the expected response of a Type-II slab is found, for $0.98 T_c>T>0.91 T_c$ the paramagnetic response in decreasing field increases when crossing the long-range vortex attraction line, and finally a pronounced paramagnetic response followed by a jump to the Meissner state is observed when crossing the $\sigma_{SN}=0$ line. Besides being useful for reaffirming our conclusions, this temperature dependence of the GPR can also be directly  verifiable experimentally. Here, the  considered samples are ideally clean, but even in realistic samples where flux trapping is present even at zero field, the rise and fall of GPR as a function of temperature will be easily observable in the above discussed scenario. Note that in general, changing any of the parameters can drive the {\it in silico} material across the crossover between the types of superconductivity, and thereby change the paramagnetic response. GPR is only sensitive on the regime of superconductivity the material is in, i.e., where the taken parameter set lies in the reconstructed Fig. \ref{critdom}(a) - Type-I, Type-II superconductivity, or in between.

\section*{Discussion}
\begin{figure}[t]
\includegraphics[width=0.6\columnwidth]{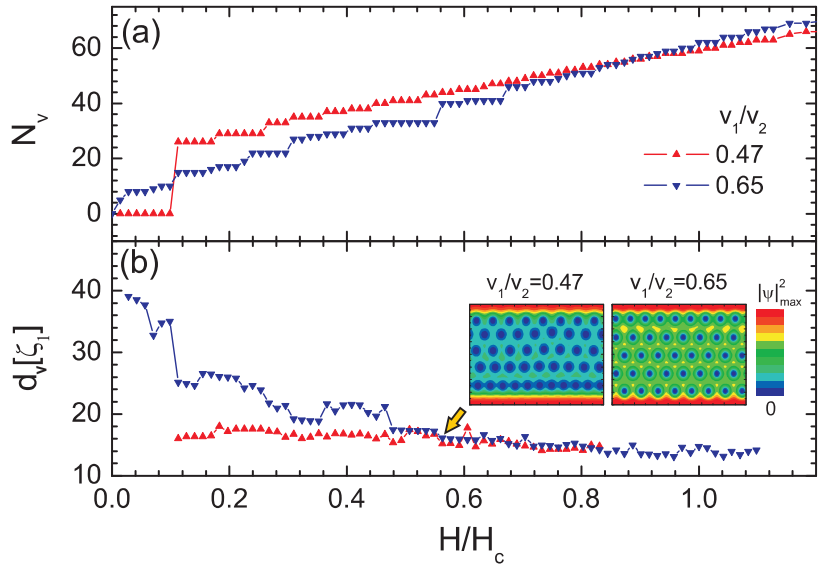}
\caption{The number of vortices $N_v$ in the sample in decreasing magnetic field (below $H=H_d$) at $T=0.94T_c$ (a), and the average distance between vortices ($d_v$), for two values of $v_1/v_2$ ratio that provide different sign of the superconducting-normal state interface energy ($\sigma_{SN}$). Insets show cumulative Cooper-pair density plots ($|\psi_1|^2$+$|\psi_2|^2$) of vortex states obtained in two considered cases for the same magnetic field $H=0.563H_c$.} \label{vortices}
\end{figure}

\begin{figure}[t]
\includegraphics[width=0.6\columnwidth]{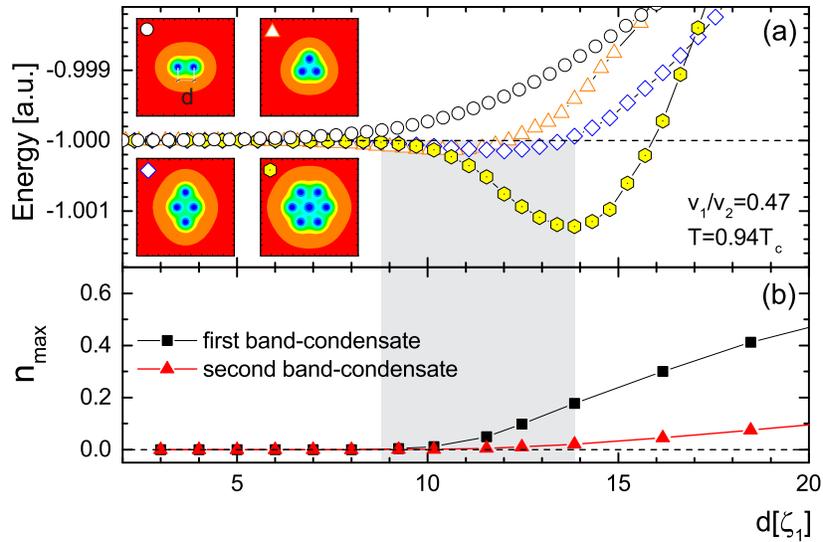}
\caption{(a) The vortex-vortex interaction energy, as a function of the distance between vortices, for parameters leading to pairwise vortex attraction ($\sigma_{SN}>0$, see open dots). Nevertheless, the short-range repulsion between vortices arises for clusters comprising more than two vortices (insets depict the cumulative Cooper-pair density distribution for the considered clusters). (b)  Maximum of the Cooper-pair density, $n_{max}$, inside the hexagonal vortex cluster for each band-condensate separately, shown as a function of vortex distance $d$ between vortices. The shaded area delimits the short-range repulsion found for the hexagonal vortex cluster.} \label{vortint}
\end{figure}

What is the underlying mechanism for the giant paramagnetic response? In simple terms, it is the facilitated trapping of magnetic flux in the crossover domain between Type-I and Type-II superconductivity, since vortices attract in the entire range of parameters where GPR is observed. However, GPR is found to be particularly large for $\sigma_{SN}>0$, where vortex-vortex interaction is purely attractive and vortices should coalesce into larger normal domains. On the contrary, we observe that in decreasing field separate vortex cores are still visible, though strongly overlapping (see inset in Fig. \ref{vortices}). 
To clarify the dense vortex packing observed in Fig. \ref{vortices}, we calculate the multibody vortex-vortex interaction shown for several vortex clusters in Fig. \ref{vortint}. As a major surprise, we found  that in this regime multibody vortex interactions become short-range repulsive and cause the formation of a vortex lattice. This is illustrated in Fig. \ref{vortint}(a) (for $v_1/v_2 = 0.47$ and $T = 0.94 T_c$, i.e. $\sigma_{SN} > 0$), where we show the calculated vortex-vortex interaction as a function of the distance between vortices (labelled d), for a vortex pair, a vortex trimer, a vortex diamond-like cluster and a hexagonal vortex cluster. The pairwise vortex interaction is purely attractive, as expected, but in the other cases the short-range repulsion arises so that energetically favorable vortex-vortex distance arises in mid-range (note that this favorable distance closely corresponds to the average vortex distance observed in Fig. \ref{vortices}(b)). An insight into the physics of this short-range repulsive interaction can be achieved by analysing the superconducting state inside, for example, the hexagonal vortex cluster shown in Fig \ref{vortint}. With this aim, we computed the maximum of the Cooper-pair density, $n_{max}$, inside that cluster for each band-condensate separately, shown as a function of vortex distance $d$ in Fig. \ref{vortint}(b). We reveal that the Cooper-pair density in the second condensate vanishes inside the vortex cluster at the vortex distance where short-range repulsion arises. Hence we can conclude that inside the vortex cluster the physics is driven by the other condensate, which has Type-II character, hence the repulsive interaction of vortices prevails at short distances. It is known that multibody vortex interactions are more complex than a simple superposition of pairwise interactions (see Refs. \cite{mbody1,mbody2,mbody3,mbody4}), but it has never been found before that multibody interactions can change the polarity of the vortex-vortex interaction. This is a key feature of the found mixed state for parameters of the system between $\sigma_{SN}=0$ and $H_c(T)=H_{c2}(T)$ lines in Fig. \ref{critdom}. In addition, we have plotted in Fig. \ref{vortices}(a) the number of vortices in the sample $N_v$  as a function of $H$ in the downward branch of $M(H)$ in Fig. \ref{mag-loop} for $v_1/v_2 = 0.65$ (in the Type-II limit) and $v_1/v_2 = 0.47$ (inside the crossover region). The high retention of flux is clearly seen as a nonlinear behavior for $v_1/v_2 = 0.47$ which contrasts the Type-II case in which $N_v$ is linearly decreasing towards the origin. We find that although the number of vortices in the states between $\sigma_{SN} = 0$ and $H_c(T) = H_{c2}(T)$ lines slowly decreases with  decreasing magnetic field, their favorable distance is approximately independent of field [see Fig. \ref{vortices}(b)]. This unconventional vortex state allows the penetration of the magnetic field in larger portions of the sample (inhomogeneous penetration, within but also between vortices), and clearly traps more flux than an ordinary vortex lattice, down to very low field - resulting in a more pronounced GPR. Due to the interlocking of vortices in this regime, the barrier for the expulsion of the entire vortex cluster in decreasing field corresponds to the Bean-Livingston barrier for a single vortex, which we confirmed by an independent calculation. Notice that as soon as the S-N surface energy changes sign, the barrier for single-vortex expulsion becomes nonzero at all fields. However, we have the simultaneous appearance of short-range vortex repulsion, which in effect diminishes the Bean-Livingston barrier and vortices are gradually expelled from the sample depending on their density and applied magnetic field. This manifests in the magnetization curves as a gradual decrease of the paramagnetic effect in decreasing field, down to zero for zero field. As the $v_1/v_2$ ratio or temperature are further increased, vortices become increasingly repulsive and the paramagnetic response decreases to its conventional behavior for Type-II superconductors.

In summary, we revealed a possibility of {\it giant paramagnetic response} in slabs of multiband superconductors (to which many recently discovered metal-borides, iron-chalcogenides, iron-pnictides, belong), with magnitude similar or multiply larger than the Meissner response for the same applied magnetic field. We showed that such unique magnetic response occurs in the crossover region between conventional types of superconductivity, and is not captured by the standard textbook descriptions. On technological end, our findings open a new class of desirable materials which can be switched to either strongly enhance or fully remove the applied magnetic field while having low power consumption. Further work is needed to characterize the behavior of these materials under e.g. applied electric current and nanostructuring or downscaling.

\section*{Methods}

In this work we used the two-component Ginzburg-Landau (TCGL) theory. In the TCGL  framework, as given in Ref. \onlinecite{Chaves2011}, eight independent material parameters are needed for a system with both interband and magnetic coupling, namely, the Fermi velocity of the first band $v_1$, the square of the ratio of the Fermi velocities in the two bands $\alpha=(\frac{v_1}{v_2})^2$, the elements of the coupling matrix $\lambda_{11}$, $\lambda_{22}$ and $\lambda_{12}=\lambda_{21}$, the total density of states $N(0)$ as well as the partial density of states of the first band $n_1$ ($n_2=1-n_1$), and finally $T_{c}$, which sets the energy scale $W^2=8\pi^2T_c^2/7\zeta(3)$. The TCGL free energy functional reads
\begin{equation}
\mathcal{F} = \sum_{j=1,2} \alpha_j |\psi_j|^2 +
\frac{1}{2}\beta_j|\psi_j|^4+\frac{1}{2m_j}|(\frac{\hbar}{i}\nabla-\frac{2e}{c}\vec{A})\psi_j|^2
-\Gamma(\psi_1^*\psi_2+\psi_1\psi_2^*) +
\frac{(\vec{h}-\vec{H})^2}{8\pi}, \label{FE}
\end{equation}
where $j=1,2$ is the band index, $\alpha_j=-N(0)n_j\chi_j=-N(0)n_j(\tau-S_j/n_j\delta)$, $\beta_j=(N(0)n_j)/W^2$, $m_j=3W^2/(N(0)n_j v_j^2)$, and $\Gamma=(N(0)\lambda_{12}) / \delta$, with $\delta$ being the determinant of the coupling matrix, and $S$, $S_1$ and $S_2$ defined as in Ref. \onlinecite{Kogan2011}. The local magnetic field in the sample is denoted by $\vec{h}$ and the external applied field by $\vec{H}$.

Minimization of the free energy in Eq. (\ref{FE}) with respect to $\psi_j$ and $\vec{A}$ yields the Ginzburg-Landau equations: two for the order parameters $\psi_1$ and $\psi_2$, and the equation for the vector potential (calculated from the supercurrent of the coupled condensate). Introducing the normalization for the order parameters by $W$, for the vector potential by $A_0=hc/4e\pi\zeta_{1}$, and for the lengths by $\zeta_{1}=\hbar v_1/\sqrt{6}W$, the dimensionless TCGL equations are written as:
\begin{eqnarray}
(-i \nabla - \vec{A})^2 \psi_1 - (\chi_1-|\psi_1|^2)\psi_1 - \gamma  \psi_2=0, \label{TDGL1}\\
\frac{1}{\alpha}(-i \nabla - \vec{A})^2 \psi_2 - (\chi_2-|\psi_2|^2)\psi_2 - \frac{\gamma \kappa_2^2}{ \alpha^2 \kappa_1^2} \psi_1=0, \label{TDGL2}\\
\kappa_1^2 \nabla \times \nabla \times \vec{A}=\vec{j}_s,\label{scurrent}
\end{eqnarray}
where $\kappa_1=\frac{3cW}{hev_1^2}\sqrt{\frac{\pi}{2n_1N(0)}}$, $\kappa_2=\kappa_1 \alpha \sqrt{n_1/n_2}$, and $\gamma=\lambda_{12}/n_1\delta$. In Eq. (\ref{scurrent}) the supercurrent density is  
\begin{equation}
 \vec{j}_s =  \mathcal{R}[\psi_1(i\nabla-\vec{A}) \psi_1^*]+ \frac{\alpha \kappa_1^2}{\kappa_2^2}\mathcal{R}[\psi_2(i\nabla-\vec{A}) \psi_2^*],
\label{scurrent2}
\end{equation}
where $\mathcal{R}$ denotes the real part of the expression. After the made choice of normalization units, we are left with six parameters: $\lambda_{11}$, $\lambda_{22}$,  $\lambda_{12}$, $v_1/v_2$, $n_1$, and $N(0)$. 

In our numerical experiment, the TCGL equations (\ref{TDGL1})-(\ref{scurrent}) were integrated self-consistently on a two dimensional grid with grid spacing $a_x=a_y=\zeta_{1}$, much smaller than any characteristic length scale at the considered temperature. The discretization was implemented by the link variable method which preserves the gauge invariance of these equations \cite{Milosevic2010}. For the iterative solver, we combined a relaxation method with a stable and accurate semi-implicit algorithm \cite{Adams2002}. Periodic boundary conditions were applied in the $x$ direction whereas for the $y$ direction we imposed Neumann boundary conditions at the superconductor-vacuum interface (for details of the numerical implementation, please see Ref. \onlinecite{Milosevic2010}). Note that due to the infinite slab geometry, the surface magnetic field equals the applied one (the demagnetizing effects are negligible), and the simulation is effectively two-dimensional (in the $(x,y)$ plane). The subsequently calculated magnetization, $M=(\langle h \rangle-H)/4\pi$ ($\langle ... \rangle$ denotes spatial averaging inside the sample), is a measure of the expelled flux from the sample and the corresponding $M(H)$ response was obtained by ramping up the magnetic field with steps of $\Delta H=2 \times 10^{-4}$ (in units of $H_0=\hbar c/2e\zeta_{1}^2$). The magnetic field is scaled to the thermodynamic critical field $H_c$, for easier comprehension of the related physics. For calculation of $H_c$ for two-band superconductors, we refer to Ref. \cite{Chaves2011}. 

\section*{Acknowledgments}
This work was supported by the Brazilian science agencies CAPES (PNPD 223038.003145/2011-00), CNPq (307552/2012-8, 141911/2012-3, and
APV-4 02937/2013-9), and FACEPE (APQ-0202-1.05/10 and BCT-0278-1.05/11), the Flemish Science Foundation (FWO-Vl), and by the CNPq-FWO
cooperation programme (CNPq 490297/2009-9). R.M.S. acknowledges support from the SRS PhD+ program of the University Cooperation for Development of the Flemish Interuniversity Council (VLIR-UOS). M.V.M. acknowledges support from CNPq (APV-4 02937/2013-9), FACEPE (APV-0034-1.05/14), and CAPES (BEX1392/11-5).


\section*{Author contributions}
R.M.S., M.V.M. and J.A.A conceived and executed the study. A.A.S. and F.M.P. contributed to discussion of the results and the final 
scientific statement of the article.

\section*{Competing financial interests}
The authors declare no competing financial interests.

\end{document}